\documentclass[]{emulateapj}
\def\pppm{\rm P^3M}
\def\mpchi{\,h^{-1}{\rm {Mpc}}}
\def\mpchii{\,h{\rm {Mpc}^{-1}}}

\def\msun{\,h^{-1}{\rm M_\sun}}
\def\k{\mathbf{k}}

\begin{document}

\title{Determine the galaxy bias factors on large scales using bispectrum method}

\author{H. Guo\altaffilmark{1,2}, Y. P. Jing\altaffilmark{1}}
\altaffiltext{1}{Key Laboratory for Research in Galaxies and
Cosmology, Shanghai Astronomical Observatory, Chinese Academy of
Sciences, Nandan Road 80, Shanghai 200030, China; guoh@shao.ac.cn,
ypjing@shao.ac.cn.} \altaffiltext{2}{Graduate School of the Chinese
Academy of Sciences, 19A, Yuquan Road, Beijing, China}

\begin{abstract}

We study whether the bias factors of galaxies can be unbiasedly
recovered from their power spectra and bispectra. We use a set of
numerical $N$-body simulations and construct large mock galaxy
catalogs based upon the semi-analytical model of \citet{Croton06}.
We measure the reduced bispectra for galaxies of different
luminosity, and determine the linear and first nonlinear bias
factors from their bispectra. We find that on large scales down to
that of the wavenumber $k=0.1\mpchii$, the bias factors $b_1$ and
$b_2$ are nearly constant, and $b_1$ obtained with the bispectrum
method agrees very well with the expected value. The nonlinear bias
factor $b_2$ is negative, except for the most luminous galaxies with
$M_r<-23$ which have a positive $b_2$. The behavior of $b_2$ of
galaxies is consistent with the $b_2$ mass dependence of their host
halos. We show that it is essential to have an accurate estimation
of the dark matter bispectrum in order to have an unbiased
measurement of $b_1$ and $b_2$. We also test the analytical approach
of incorporating halo occupation distribution to model the galaxy
power spectrum and bispectrum. The halo model predictions do not fit
the simulation results well on the precision requirement of current
cosmological studies.

\end{abstract}


\section{Introduction}

In the current $\Lambda$CDM cosmological scenario, dark matter
dominates the matter density in the universe, and luminous galaxies
are thought to be formed within dark matter halos. However, detailed
physical processes of galaxy formation, such as star formation and
supernova feedback, are very complicated. Different prescriptions of
galaxy formation may lead to different relations between luminous
galaxies and the underlying dark matter, or equivalently, different
galaxy biases (e.g., \citealt{Kaiser84,Davis85,Bardeen86}).
Therefore, the key to our understanding of galaxy formation is to
study the galaxy bias both in theories and observations. If galaxy
formation is mainly determined by local physical processes (such as
hydrodynamics), the galaxy bias is then a constant on large scales
\citep{coles93}. In the first order, the galaxy density fluctuation
is proportional to that of the dark matter on large scales,
$\delta_g \propto \delta_m$. The coefficient here is usually called
as the linear bias factor, $b_1$. It is important to study its
dependence on the luminosity, morphology, mass, and redshift of
galaxies (e.g.,
\citealt{Xia87,boerner91,Norberg02,Jing02,Zehavi04,Park05,Li06,Meneux08}),
because this information provides important clues to how galaxies
are formed.

The bias factor $b_1$ and its relation with the scale $k$ can be
determined directly from the ratios between the power spectra of
galaxies ($P_g(k)$) and those of the underlying dark matter
($P_m(k)$),

%%%%%%%%%%%%%%%%%%%%%%%%%%%%%%%%%%%%%%
\begin{equation}
P_g(k) = b_1^2(k) P_m(k), \label{eqn:b1p}
\end{equation}
%%%%%%%%%%%%%%%%%%%%%%%%%%%%%%%%%%%%%%
where the higher-order terms are negligible. But it is important to
note that the power spectrum of dark matter is usually not known in
observations. Therefore, the absolute value of $b_1(k)$ cannot be
obtained through the above relation, unless the power spectrum of
dark matter is measured from observations, such as future weak
lensing surveys. From Equation (\ref{eqn:b1p}), we note that the
bias factor $b_1$ actually couples with the properties of the
large-scale matter density field, such as the linear rms
fluctuation, $\sigma_8$. And it is impossible to break such
degeneracy only by using the galaxy power spectrum. It is then
challenging to determine the galaxy bias purely in observations.

Fortunately, there is a way of doing this by involving the
second-order statistics, the bispectrum $B(k_1,k_2,k_3)$, or more
conveniently the reduced bispectrum $Q(k_1,k_2,k_3)$, which is
defined by

%%%%%%%%%%%%%%%%%%%%%%%%%%%%%%%%%%%%%%%%%%%%%%%%%%
\begin{equation}
Q(k_1,k_2,k_3)=\frac{B(k_1,k_2,k_3)}{P(k_1)P(k_2)+cyc},\label{eqn:qbp}
\end{equation}
%%%%%%%%%%%%%%%%%%%%%%%%%%%%%%%%%%%%%%%%%%%%%%%%%%
where ($k_1,k_2,k_3$) are the three sides of a triangle and $P(k)$
is the corresponding power spectrum (galaxy or dark matter). On
sufficiently large scales where the density fluctuation is small, we
can expand the galaxy density field $\delta_g$ in the Taylor series
of $\delta_m$, which up to the second order can be written as

%%%%%%%%%%%%%%%%%%%%%%%%%%%%%%%%%%%%%%
\begin{equation}
\delta_g=b_1\delta_m+\frac{1}{2}b_2\delta_m^2,\label{eqn:taylor}
\end{equation}
%%%%%%%%%%%%%%%%%%%%%%%%%%%%%%%%%%%%%%
where $b_1$ is the linear bias factor and $b_2$ is the first
nonlinear bias factor. If we assume that the bias factors are
deterministic, i.e., constant $b_1$ and $b_2$, we can derive the
following relation from Equation (\ref{eqn:taylor}) by utilizing the
reduced bispectra $Q_g$ of galaxies and $Q_m$ of dark matter(e.g.,
\citealt{Fry94,Gaztanaga94,Mo97}),

%%%%%%%%%%%%%%%%%%%%%%%%%%%%%%%%%%%%%%
\begin{equation}
Q_g(k_1,k_2,k_3)=\frac{Q_m(k_1,k_2,k_3)}{b_1}+\frac{b_2}{b_1^2}.
\label{eqn:bfit}
\end{equation}
%%%%%%%%%%%%%%%%%%%%%%%%%%%%%%%%%%%%%%
According to the second-order perturbation theory, the reduced
bispectrum of dark matter $Q_m(k_1,k_2,k_3)$ can be predicted on
large scales based on the measurement of galaxy power spectrum. In
the second-order perturbation theory(hereafter PT2), it gives
\citep{Fry84,Matarrese97,Bernardeau02}

%%%%%%%%%%%%%%%%%%%%%%%%%%%%%%%%%%%%%%%%%%%%%%%%%%
\begin{eqnarray}
B_{PT}(k_1,k_2,k_3) &=& F(\mathbf{k_1},\mathbf{k_2})
P_L(k_1)P_L(k_2)+ cyc \label{eqn:Bk}
\\
F(\mathbf{k_1},\mathbf{k_2}) &=& (1 + \mu) + \frac{\mathbf{k_1}
\cdot \mathbf{k_2}}{k_1k_2}\left(\frac{k_1}{k_2} +
\frac{k_2}{k_1}\right)
\nonumber\\
&&+ (1 - \mu)\left(\frac{\mathbf{k_1} \cdot
\mathbf{k_2}}{k_1k_2}\right)^2,
\end{eqnarray}
%%%%%%%%%%%%%%%%%%%%%%%%%%%%%%%%%%%%%%%%%%%%%%%%%%
where $\mu=3\Omega_m^{-2/63}/7$ denotes the mild dependence on
cosmology and $P_L(k)$ is the linear dark matter power spectrum. It
is easy to see that within the tree-level PT, the reduced bispectrum
$Q_m(k_1,k_2,k_3)$ does not depend on redshift or amplitude of the
matter density fluctuation, but weakly depends on the cosmology,
thus depends almost solely on the shape of the matter power
spectrum. Therefore, the reduced bispectrum $Q_m(k_1,k_2,k_3)$, on
large scales, can be predicted from the measurement of galaxy power
spectrum, since the galaxy power spectrum $P_g(k)$ has the same
shape as the matter power spectrum $P_m(k)$ (they are proportional
to each other on large scales). The bias factors $b_1$ and $b_2$ can
then be determined from the measurement of $Q_g$ and the prediction
of $Q_m$, with at least two independent triangle configurations from
the galaxy distribution. Many more configurations are actually used
in the real measurement, because the measurement errors are present.
With the value of $b_1$, one can determine the amplitude of the
linear density fluctuation from the galaxy distribution alone, which
can thus become a potentially powerful way for dark energy study if
the density fluctuation amplitude can be measured in this way for a
few redshifts. The values of $b_1$ and $b_2$ also contain very
important information about the formation of galaxies.

With this method, \cite{Verde02} measured the galaxy bias parameters
using the observational data of 2dF Galaxy Redshift Survey(2dFGRS).
They found that the linear bias factor $b_1$ is consistent with
unity and the quadratic bias parameter $b_2$ is consistent with
zero. Though the error bars of their measurements are also large,
the results may still indicate that galaxies follow the distribution
of dark matter. The range of scales assumed, however, was in the
quasilinear and nonlinear regime ($0.1 < k < 0.5\mpchii$). On these
scales, the second-order perturbation theory is already not accurate
enough to predict the dark matter bispectrum, and even the bias
expansion (Equation (\ref{eqn:bfit})) may have already broken down,
as we will show below.

Considering the potential applications of this method to constrain
the cosmological parameters (e.g., the amplitude of the dark matter
density distribution, the equation of state of dark energy) from the
measurement of $b_1$, and to constrain theories of galaxy formation
from measuring $b_1$ and $b_2$, it is important to test whether and
on which scale this method can produce an unbiased measurement of
$b_1$ and $b_2$. Because the bias factors are generally dependent on
the scale $k$(against the assumption of scale-invariant $b_1$ and
$b_2$), we need to check the range of validity of Equation
(\ref{eqn:bfit}) and at the same time, we have to make sure whether
it is consistent with Equation (\ref{eqn:b1p}). We should keep in
mind that for this method to be valid, the prediction of $Q_m$ must
be accurate(i.e., PT2 must be precise on large scales) and the
number density of galaxies can be expanded as in Equations
(\ref{eqn:taylor}) and (\ref{eqn:bfit}).

In our companion paper\citep{gj09}, we have studied the accuracy of
PT2 for the bispectrum of dark matter and we found that the PT2
prediction is not very accurate even on scales $k\approx
0.1\mpchii$, and high-order loop corrections are needed
\citep[e.g.,][]{Scoccimarro98,Bernardeau08}. In the current paper,
we will focus on the second point, i.e., on which scale $\delta_g$
can be expanded as in Equation (\ref{eqn:taylor}). We generate a set
of catalogs of mock galaxies from a semi-analytical model of galaxy
formation. Then we measure $b_1$ and $b_2$ from the galaxy reduced
bispectrum. The obtained linear bias factor $b_1$ is compared with
the expected value determined by the ratio of power spectra
(Equation (\ref{eqn:b1p})) in the numerical simulations. We will
also demonstrate that an accurate prediction of the dark matter
bispectrum is necessary for obtaining correct bias factors. We
constrain our discussion only in the real space(relative to the
redshift space) to avoid the complicated observational effects.

The paper is constructed as follows. We describe our simulations and
galaxy mock catalogs in Section \ref{simulations}. The accuracy of
determining the galaxy bias factors using bispectrum is displayed in
Section \ref{galaxybias}. We use the halo model to study the
influence of different components in Section \ref{halomodel}. We
summarize our results in Section \ref{conclusions}.

\section{$N$-body Simulations}
\label{simulations}

The cosmology considered here is a canonical spatially flat cold
dark matter model with the density parameter $\Omega_m=0.268$, the
cosmological constant $\Omega_\Lambda=0.732$, the Hubble constant
$h=0.71$, and the baryon density parameter $\Omega_b=0.045$. The
primordial density field is assumed to be Gaussian with a
scale-invariant power spectrum $\propto$$k$. For the linear power
spectrum, we generate it with the CMBfast code \citep{Seljak96}, and
the normalization is set to $\sigma_8=0.85$, where $\sigma_8$ is the
present linear rms density fluctuation within a sphere of radius
$8\mpchi$.

\begin{deluxetable}{cccccc}
\tablewidth{0pt} \tablecolumns{4} \tablecaption{Simulation
parameters} \tablehead{\colhead{boxsize($\mpchi$)} & \colhead
{particles} & \colhead {realizations} & \colhead {$m_{\rm
particle}$}} \startdata
600 & $1024^3$ & 3 & $1.5\times 10^{10}h^{-1}M_\odot$\\
1200 & $1024^3$ & 4 & $1.2\times 10^{11}h^{-1}M_\odot$
\enddata
\label{tab:simu}
\end{deluxetable}

We use an upgraded version of the Particle-Particle-Particle-Mesh
($\pppm$) code of \citet{js98,js02} to simulate structure formation
in the universe. The code has now incorporated the multiple level
$\pppm$ gravity solver for high-density regions \citep{js00}. In
order to have a large mass resolution range, we run a total of $7$
simulations with $1024^3$ particles in different simulation boxes,
which we hereafter denote by $L600$ and $L1200$ by different box
sizes(Table \ref{tab:simu}) \citep{Jing07}. The simulations were run
on an SGI Altix 350 with 16 processors with OPENMP parallelization
in Shanghai Astronomical Observatory. Dark matter halos are
identified using the standard Friends-of-Friends(FOF) algorithm with
a linking length $b$ equal to $0.2$ times the mean particle
separation. Unbound particles (with positive binding energy) are
excluded to avoid contamination. The different resolutions of the
simulations enable us to check the consistency among the results
from simulations of different $L_{box}$, as well as to investigate
the behavior of bispectrum over a large dynamical range. Here, we
choose the Fourier space bin scheme as $\Delta log_{10}(k)=0.1$ for
$k<0.1\mpchii$ and $0.05$ for $k>0.1\mpchii$ when measuring the
bispectrum.

For the galaxy samples, we build our catalogs on the semi-analytical
model of \citet{Croton06}. They have generated the galaxy
distribution in the Millennium Simulation with the box size
$L_{box}=500\mpchi$ on a side\citep{Springel05}. The cosmological
parameters used in \cite{Croton06}($\Omega_m=0.25$,
$\Omega_\Lambda=0.75$, $\Omega_b=0.045$, $h=0.73$, $\sigma_8=0.9$)
are not exactly the same as ours. But the slight difference in the
parameters is not the focus in this paper, since an approximate
relation connecting galaxies to dark matter is sufficient here. For
each dark matter halo in our simulations, we randomly choose a
corresponding galaxy host halo of the same mass from the galaxy
catalog of \citet{Croton06}. One thing to note is that the virial
masses of the halos in \cite{Croton06} are defined by the spherical
overdensity approach, which makes their final halo masses about
$1.5$ times smaller than those of our corresponding FOF halos. We
adjust the masses of \cite{Croton06} halos, so they would match our
halo definitions. After the adjustment, we put those galaxies of the
host halos of \cite{Croton06} into our corresponding dark matter
halos, while preserving the same relative positions to the halo
centers. Finally, we obtain the corresponding distributions of
galaxies in our simulations, as well as their properties, such as
the luminosity and velocity. We use the r-band magnitude for the
luminosity of galaxies, $M_r$, which is defined to be
$M_r=M_{abs}+5\lg h$, that is, the absolute magnitude $M_{abs}$ when
$h=1$.

\begin{figure}
\epsscale{1.2}
\plotone{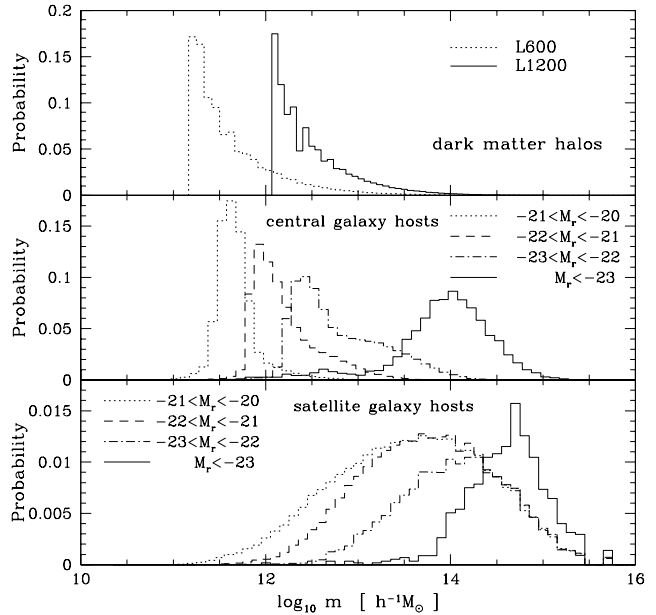} \caption{Probability distribution of
the virial mass $m$ of the dark matter halos in our simulations
(Top). The host halo mass distributions of the central and satellite
galaxies in the galaxy catalog of \citet{Croton06} are also shown in
the middle and bottom panels.} \label{fig:pdfmh}
\end{figure}

We show in the top panel of Figure \ref{fig:pdfmh} the probability
distribution of the virial mass $m$ of our simulation dark matter
halos. $L1200$ has more massive halos than $L600$, thus $L1200$ can
be used to investigate the distribution of more luminous galaxies
while $L600$ for ordinary galaxies with moderate luminosity. The
sharp cut for each $L_{box}$ represents the minimum mass of the dark
matter halos that consist of at least $10$ particles in our
definition. The host halo mass distributions of the central and
satellite galaxies in the galaxy catalog of \citet{Croton06} are
also shown in the middle and bottom panels of Figure
\ref{fig:pdfmh}, respectively. We divide the galaxy samples into
subsamples of different luminosity, and the probability distribution
of host halo masses are shown as the different lines in the figure.

To avoid incompleteness, the available galaxy sub-samples are
actually limited for each $L_{box}$ simulation. As it is shown in
Figure \ref{fig:pdfmh}, we can use all those galaxies with $M_r<-20$
for $L600$ (However, the most luminous galaxies with $M_r<-23$ are
not abundant in $L600$, which will cause large fluctuations in their
bispectrum statistics. That is why we do not consider this
luminosity subsample for $L600$). For $L1200$, because dark matter
halos have high mass of $m>10^{12}\msun$, only those galaxies of
$M_r<-22$ will be analyzed.

\section{Galaxy biasing}
\label{galaxybias}

To obtain the galaxy bias factors by fitting Equation
(\ref{eqn:bfit}), one needs an accurate error estimation for the
galaxy bispectrum $Q_g$. Since we do not have enough independent
Fourier space k modes on large scales of $k<0.1\mpchii$
(finite-volume effect), and the number of realizations for each
$L_{box}$ simulation are also limited, the large fluctuations among
different realizations on large scales would prevent us from making
an accurate error estimation of $Q_g$. So we use the Gaussian
uncertainties, instead of the simulation fluctuations, as the errors
for $Q_g$.

The uncertainty of bispectrum for a Gaussian density field reads
(e.g., \citealt{Fry93,Scoccimarro98,Scoccimarro04,Sefusatti07})

%%%%%%%%%%%%%%%%%%%%%%%%%%%%%%%%%%%%%%%%%%%%%%%%%%
\begin{eqnarray}
\label{eqn:berr} \langle\Delta B^2_{g}\rangle &=& \frac{1}{N_{123}}\
P_{tot}(k_1)P_{tot}(k_2)P_{tot}(k_3) \\
P_{tot}(k) &=& P_g(k)+\frac{1}{N_p},
\end{eqnarray}
%%%%%%%%%%%%%%%%%%%%%%%%%%%%%%%%%%%%%%%%%%%%%%%%%%
where $N_{123}$ is the number of independent triangle configuration
modes in the Fourier space, and $P_{tot}(k)$ includes the shot noise
in the galaxy power spectrum for the case of $N_p$ objects. Although
we measure $N_{123}$ directly from the simulations, it can be
estimated theoretically as \citep{Sefusatti07}

%%%%%%%%%%%%%%%%%%%%%%%%%%%%%%%%%%%%%%%%%%%%%%%%%%
\begin{equation}
N_{123}\approx \frac{8\pi^2}{s_{123}k_f^6}k_1k_2k_3\Delta k_1 \Delta
k_2 \Delta k_3,
\end{equation}
%%%%%%%%%%%%%%%%%%%%%%%%%%%%%%%%%%%%%%%%%%%%%%%%%%
where $k_f=2\pi/L_{box}$ and $s_{123}=6,2,1$ for equilateral,
isosceles and general triangles. By assuming that the variance of
bispectrum dominates over that of the power spectrum, variance of
the reduced bispectrum is then given by,

%%%%%%%%%%%%%%%%%%%%%%%%%%%%%%%%%%%%%%%%%%%%%%%%%%
\begin{equation}
\langle\Delta Q^2_g\rangle = \frac{\langle\Delta
B^2_{g}\rangle}{[P_g(k_1)P_g(k_2)+ cyc]^2}. \label{eqn:gerr}
\end{equation}
%%%%%%%%%%%%%%%%%%%%%%%%%%%%%%%%%%%%%%%%%%%%%%%%%%
The fluctuation of $Q_g$ is thus directly determined by the number
of independent triangle modes, $N_{123}$, which are actually related
to the simulation box size, $L_{box}$. The larger box simulation has
higher precision in the determination of bispectrum on large scales,
especially for $k<0.1\mpchii$. In addition to the considerable
fluctuations, the finite-volume effect is also significant for the
bispectrum on large scales. Because the bispectrum essentially
reflects the influence of the gravitational instability, the
existence of large-scale structures, such as filaments, will
significantly affect the final bispectrum estimation(e.g.,
\citealt{Sefusatti05}). Overall, large $L_{box}$ simulations are
essential if we want to determine the bispectrum on such large
scales with unerring accuracy, as we have processed in this paper.

\begin{figure}
\epsscale{1.2}
\plotone{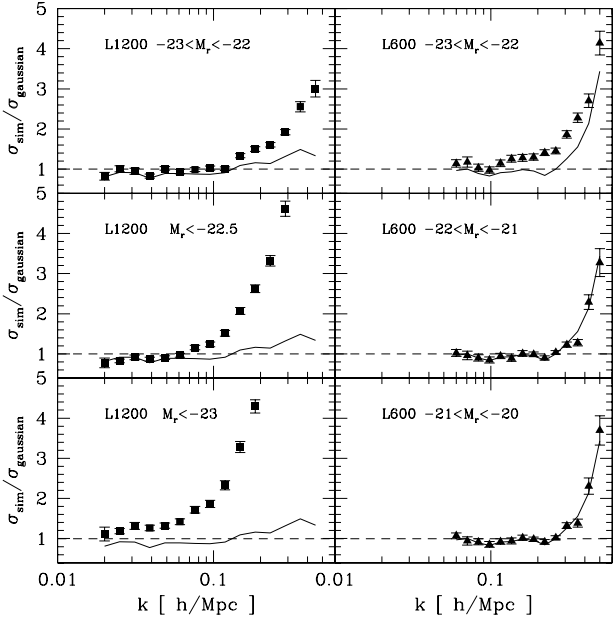} \caption{Ratios of the reduced
bispectrum errors measured from simulations to the errors based on
the Gaussian field assumption. Those points are for galaxies of
different luminosity (as shown in each panel), while the solid lines
represent the ratios for dark matter.} \label{fig:sigrg}
\end{figure}

Figure \ref{fig:sigrg} shows the ratios between the errors on the
mean of $Q$ averaged over the realizations and the corresponding
uncertainties based on the Gaussian assumption. The ratios for
galaxies of different luminosity (symbols) are compared with those
ratios for dark matter (solid lines) for both $L1200$ and $L600$.
Given that triangles with a side in common will correlate with each
other, we alleviate this problem by defining the scale $k$ in Figure
\ref{fig:sigrg} to be the longest side of the triangle and by
including all the triangle configurations in each $k$ bin. The
remaining correlations are ignored. So, each data point in Figure 2
represents the mean of all the triangle configurations whose longest
sides are in the same k bin, and the $1-\sigma$ error shown is the
fluctuation among the different triangle configurations. The small
discrepancies of dark matter ratios between $L1200$ and $L600$ are
caused by the different simulation resolutions, with $L1200$ more
reliable on large scales and $L600$ on small scales. As for galaxies
of $M_r>-23$, the reduced bispectrum errors can be well described by
the Gaussian errors on large scales of $k<0.1\mpchii$. The galaxies
have similar error ratios to those of dark matter, implying that the
deviation of the bispectrum error from the Gaussian field error is
slim in the linear clustering regime.  For the galaxy samples of
$M_r<-23$, the simulation variances deviate rapidly from the
Gaussian uncertainties even on large scales. This is because these
galaxies are highly biased, and the non-Gaussianity already plays an
important role in determining $\langle\Delta Q^2_g\rangle$ for such
luminous galaxies. Since we are interested in the behavior of the
bias factors at $k<0.2 \mpchii$, we will use Equation
(\ref{eqn:gerr}) to estimate the errors for the galaxy reduced
bispectrum.

\begin{figure}
\epsscale{1.2}
\plotone{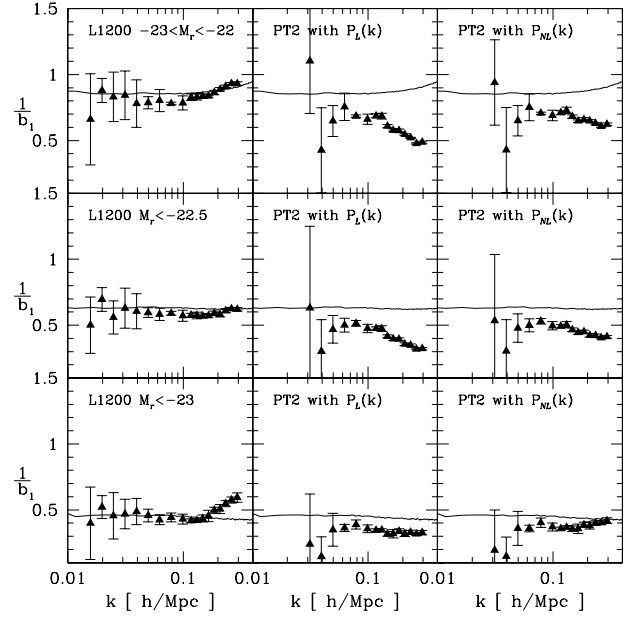} \caption{$\chi^2$ fitting results of
Equation (\ref{eqn:bfit}) for the linear bias factor $b_1(k)$ in the
galaxy samples of $L1200$. The left panels show the fittings with
$Q_m$ determined from $L1200$ simulations. In the middle panels,
$Q_m$ is estimated with PT2 using the linear power spectrum $P_L(k)$
in Equation (\ref{eqn:qbp}) and Equation (\ref{eqn:Bk}). In the
right panels, $Q_m$ is also estimated with PT2, but the nonlinear
power spectrum $P_{NL}(k)$ estimated directly from the simulation is
used for the PT2 predictions. The solid lines denote the expected
value derived from the power spectrum ratios in Equation
(\ref{eqn:b1p}). The scale $k$ is defined as the maximum value of
the triangle sides $(k_1,k_2,k_3)$.} \label{fig:b112}
\end{figure}

\begin{figure}
\epsscale{1.2}
\plotone{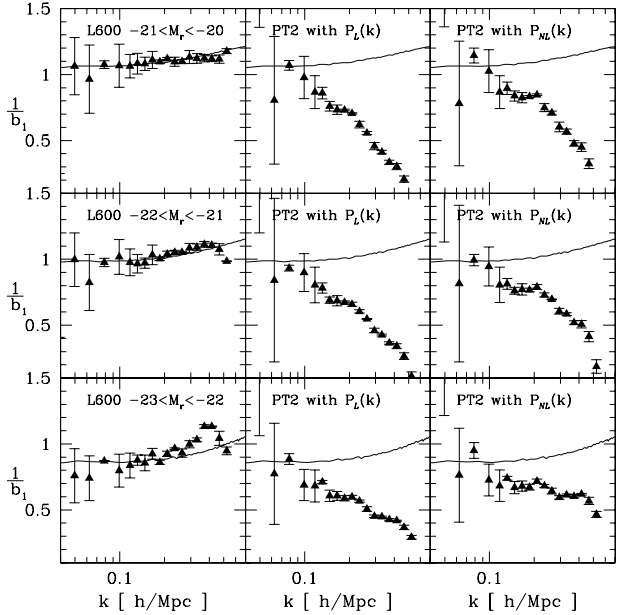} \caption{Same as Figure
\ref{fig:b112}, but for $L600$ galaxy samples.} \label{fig:b16}
\end{figure}

In Figure \ref{fig:b112} ($L1200$) and Figure \ref{fig:b16}
($L600$), we show the $\chi^2$ fitting results of Equation
(\ref{eqn:bfit}) for the linear bias factor $b_1(k)$. The left
panels show the fitting results with $Q_m$ measured from
simulations, and the middle panels show the results with $Q_m$
determined by PT2. As in observations, we also use the nonlinear
power spectrum $P_{NL}(k)$ in Equations (\ref{eqn:qbp}) and
(\ref{eqn:Bk}) for the PT2 estimation of $Q_m$, and the results are
shown in the right panels. The solid lines denote the predictions
derived from the power spectrum ratios as in Equation
(\ref{eqn:b1p}). Again, the scale $k$ in the figures is defined as
the maximum triangle size of ($k_1,k_2,k_3$). And for each $k$ bin,
we have included all types of triangle shapes to make the fittings.
The jump in the errors of $L600$ at $0.1\mpchii$ is caused by our
change of Fourier space bin scheme at this scale.

In the left panels, the bias factor $b_1(k)$ determined by Equation
(\ref{eqn:bfit}) is well consistent with that of the power spectrum
ratio (Equation(\ref{eqn:b1p})) on large scales of $k<0.15\mpchii$.
This indicates that $b_1(k)$ can be reconstructed from bispectrum
observations of galaxies without significant systematic bias.
However, when $Q_m$ is calculated from PT2 (the middle panels), the
fitted $b_1(k)$ are overestimated on most scales except for the
largest one of $k\sim0.03\mpchii$ where the fluctuation is dramatic.
The main reason is that PT2 overestimates $Q_m$ for colinear
configurations ($\alpha\equiv arccos(\k_1 \cdot \k_2)/k_1 k_2=0$ and
$\pi$) and underestimates for triangles of $\alpha\approx 0.6\pi$
even on large scales of $k \approx 0.1\mpchii$ \citep[][]{gj09}.
Using the nonlinear power spectrum for the PT2 prediction of $Q_m$
does not improve the fitting results, as shown by the right panels.
Our results clearly show that it is necessary to have an accurate
estimation of the dark matter bispectrum if Equation
(\ref{eqn:bfit}) is used to get the galaxy bias factors.

When the scale goes down to $k>0.15\mpchii$, the nonlinear effects
become important. The linear bias $b_1(k)$ does not keep a constant
on these small scales. Then the premise of the fittings that $b_1$
and $b_2$ both are scale-invariant in Equation  (\ref{eqn:bfit})
breaks down. We would expect that these bias factors from the power
spectrum ratios and from the bispectrum fittings will be different,
though we note that they agree well on scales down to $k=0.4\mpchii$
for galaxy samples of $M_r > -23$, which could be a coincidence. As
PT2 breaks down for the prediction of dark matter bispectrum even on
scales slightly larger than that of $k=0.1\mpchii$, our results show
the expansion of the galaxy density fields in Equation
(\ref{eqn:taylor}) is valid on the smaller scales ($k=0.15\mpchii$).

\begin{figure}
\epsscale{1.2}
\plotone{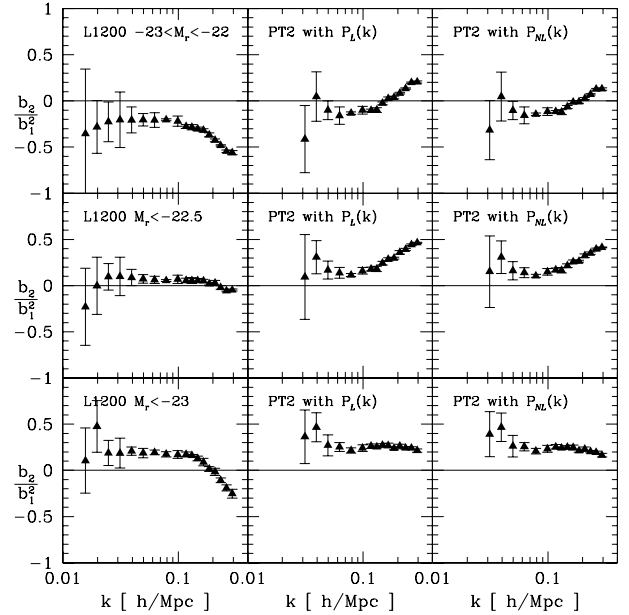} \caption{Same as Figure \ref{fig:b112}, but for the
scale dependence of $b_2/b_1^2$. The solid lines show the case of a
vanishing nonlinear bias factor $b_2$ for comparison. Since the
error bars are much too large for $k<0.03\mpchii$ in the middle and
right panels, we do not show those data points here.}
\label{fig:b212}
\end{figure}

\begin{figure}
\epsscale{1.2}
\plotone{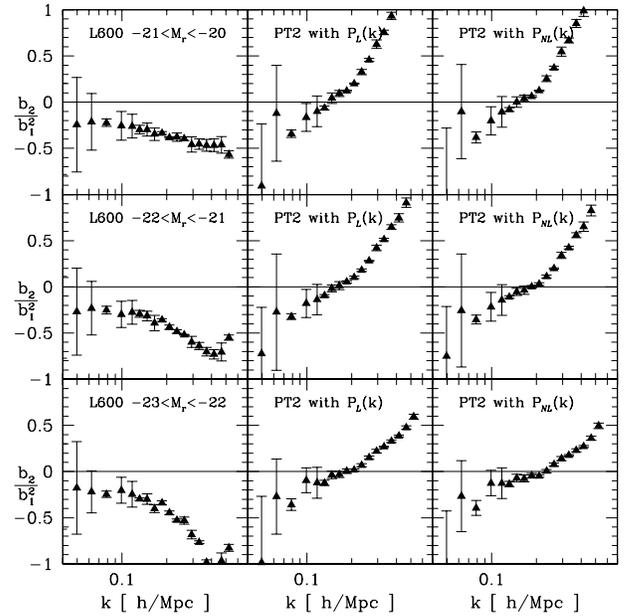} \caption{Same as Figure
\ref{fig:b212}, but for $L600$ simulations.} \label{fig:b26}
\end{figure}

To verify the conclusions above, we also show the scale dependence
of $b_2/b_1^2$ in Figure \ref{fig:b212} ($L1200$) and Figure
\ref{fig:b26} ($L600$). We see that $b_2/b_1^2$ is nearly a constant
on large scales of $k<0.15\mpchii$. With the scale independence of
$b_1$ on these scales, we infer that the nonlinear bias $b_2$ is
also scale invariant for small $k$. It again confirms the
feasibility of deriving $b_1$ and $b_2$ from Equation
(\ref{eqn:bfit}) through direct $\chi^2$ fittings. But $b_2$
deviates from such constancy as the scale goes to the nonlinear
scales, similar to the case of the linear bias $b_1$. In general,
the nonlinear bias $b_2$ is nonvanishing on large scales and plays
an important role in the determination of the linear bias factor
$b_1$ when using Equation (\ref{eqn:bfit}). It is interesting to
note that the most luminous samples have positive values of $b_2$
($b_2\approx 1.2$ for $M_r<-23$ and $b_2\approx 0.3$ $M_r<-22.5$) on
large scales, but fainter samples have negative values of $b_2$
(e.g., $b_2\approx -0.2$ for $-21<M_r<-20$ and $b_2\approx -0.3$ for
$-23<M_r<-22$). We note that the values of $b_2/b_1^2$ for galaxy
samples with $M_r>-22.5$ are nearly the same with $b_2/b_1^2 = -0.2$
at $k<0.1\mpchii$, which might be a coincidence for this particular
semi-analytical model of galaxy formation. The general behavior that
$b_2$ increases with the luminosity at the high luminosity end, and
decreases with increasing luminosity at fainter luminosity, is
consistent with the change of nonlinear halo bias $b_2(m)$ with the
dark matter halo mass $m$ \citep{Mo97}.

The middle and right panels of Figure \ref{fig:b212} and Figure
\ref{fig:b26} clearly demonstrate the failure of the fittings for
the bias factors when using PT2 to calculate $Q_m$. (Since the error
bars are much too large for $k<0.03\mpchii$, we do not show those
data points here.) Thus PT2 is actually not a good estimator for the
reduced dark matter bispectrum $Q_m$ when we want to use it to fit
the bias factors with Equation (\ref{eqn:bfit}). \cite{Verde02} used
PT2 as the preferred choice for the estimation of $Q_m$ and they
determine the bias factors for $k=0.1 \sim 0.5\mpchii$. As we showed
above, on these scales, the PT2 is not accurate
enough\citep[see][]{gj09} and the fitting results of the bias
factors are biased even if the measurement of $Q_g$ from galaxy
surveys is accurate.

\section{An attempt to study the bispectrum at quasilinear regime with the halo  occupation model}
\label{halomodel} Since the PT2 as well as the Taylor expansion
breaks down quickly on quasilinear scales ($k>0.1\mpchii$), we want
to explore if the halo occupation distribution(HOD) model of
galaxies \citep[e.g.,][]{Jing982,Berlind02,Yang03,Zheng05} can
extend the theoretical modeling to quasilinear and even nonlinear
scales. The HOD model has been used to predict the bispectrum of
galaxies in numerical simulations
\citep[][]{Jing98,Jing04,Nichol06,Kulkarni07,Nishimichi07,Marin08}
and in analytical modeling
\citep[][]{Scoccimarro01,Takada03,Wang04}. In this section, we check
if the bispectra of galaxies on quasilinear scales can be modeled
with the latter method.

The necessary halo model ingredients are described as follows. The
dark matter halos are defined as objects with a mean density
$\Delta_{vir}$ times that of the background universe\citep{Bryan98}
where $\Delta_{vir} \approx 361$ for our cosmology parameters, and
their density distributions follow the Navarro--Frenk--White (NFW)
profile \citep{Navarro97}. The concentration parameter $c(m)$ of the
halos is given by the relation $c(m)=c_0(m/m_*)^{\beta}$, where
$c_0=9$, $\beta=-0.13$, and $m_*=4.8\times10^{12}h^{-1}M_{\odot}$ is
the nonlinear mass scale \citep{Bullock01}. We also use the same
linear power spectrum as that used for generating the initial
condition for the simulations. For the halo mass function(MF)
$n(m)$, we consider the analytical models of \cite{Press74}(PS) and
\cite{Sheth99}(ST). For the halo bias parameters $b_i(m)(i=1,2)$, we
use the corresponding results of \cite{Mo97} and \cite{Sheth99} for
PS and ST mass functions, respectively. The last ingredient of the
model are the different moments of the galaxy distribution within
the parent halos, i.e., the halo occupation number $\langle
N_g(m)\rangle$, the second moment $\langle N_g(N_g-1)\rangle$ and
the third moment $\langle N_g(N_g-1)(N_g-2)\rangle$. They are
derived directly from the galaxy mock catalogs, and are inserted
into the integrals of the halo model with interpolations of the
simulation data points.

Following \cite{Scoccimarro01}, the galaxy power spectrum and
bispectrum are given by

%%%%%%%%%%%%%%%%%%%%%%%%%%%%%%%%%%%%%%%%%%%%%%%%%%
\begin{eqnarray}
P_g(k) &=& \left[ G_{11}(k)\right]^2 P_L(k)+G_{02}(k,k) \\
B_g(k_1,k_2,k_3) &=&
G_{11}(k_1)G_{11}(k_2)G_{11}(k_3)B_{PT}\nonumber\\
&+&\left[ G_{11}(k_1)G_{11}(k_2)G_{21}(k_3)P_L(k_1)P_L(k_2) + {\rm
cyc} \right]\nonumber \\
&+&\left[ G_{11}(k_1)G_{12}(k_2, k_3)P_L(k_1)+{\rm cyc}\right]
\nonumber \\
&+& G_{03}( k_1, k_2, k_3),
\end{eqnarray}
%%%%%%%%%%%%%%%%%%%%%%%%%%%%%%%%%%%%%%%%%%%%%%%%%%
where

%%%%%%%%%%%%%%%%%%%%%%%%%%%%%%%%%%%%%%%%%%%%%%%%%%
\begin{eqnarray}
G_{ij}(k_1,\ldots,k_j) &\equiv& \int dm n(m) \frac{\langle
N_g^j(m)\rangle}{\bar{n}_g^j} b_i(m) \nonumber\\
&& \times [u(k_1|m) \ldots u(k_j|m)]
\end{eqnarray}
%%%%%%%%%%%%%%%%%%%%%%%%%%%%%%%%%%%%%%%%%%%%%%%%%%
and $b_0\equiv 1$. $\langle N_g^j(m)\rangle$ represents the
different moments of the galaxy distribution mentioned above,
$u(k|m)$ is the normalized Fourier transform of the dark matter halo
density profile $\rho(r|m)$ truncated at the virial radius, and
$\bar{n}_g$ denotes the mean number density of galaxies,

%%%%%%%%%%%%%%%%%%%%%%%%%%%%%%%%%%%%%%%%%%%%%%%%%%
\begin{equation}
\bar{n}_g = \int dm n(m)\langle N_g(m)\rangle.
\end{equation}
%%%%%%%%%%%%%%%%%%%%%%%%%%%%%%%%%%%%%%%%%%%%%%%%%%
We note that $\langle N_g^j(m)\rangle$ can be further decomposed
into the central and satellite galaxy components and for the central
galaxies, $u(k|m)=1$. The galaxy reduced bispectrum $Q_g$ is then
defined as

%%%%%%%%%%%%%%%%%%%%%%%%%%%%%%%%%%%%%%%%%%%%%%%%%%
\begin{equation}
Q_g(k_1,k_2,k_3)=\frac{B_g(k_1,k_2,k_3)}{P_g(k_1)P_g(k_2)+cyc}.
\end{equation}
%%%%%%%%%%%%%%%%%%%%%%%%%%%%%%%%%%%%%%%%%%%%%%%%%%

\begin{figure}
\epsscale{1.2}
\plotone{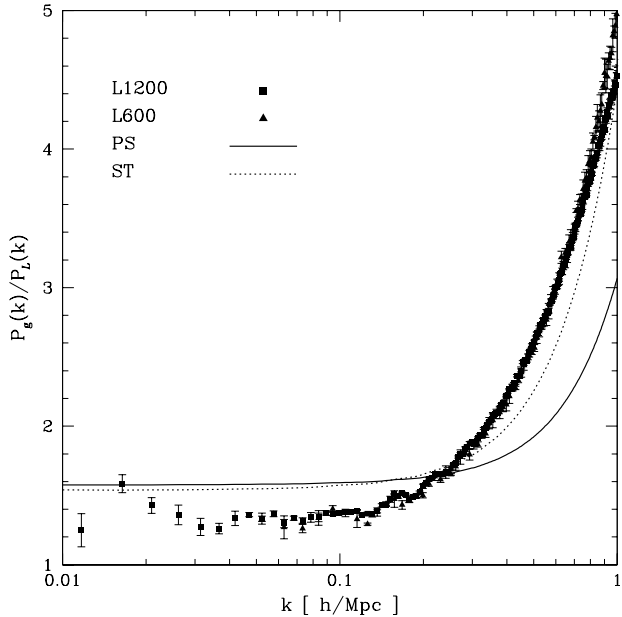} \caption{Ratio of the galaxy power
spectrum $P_g(k)$ to the linear dark matter power spectrum $P_L(k)$
for the galaxy samples of $-23<M_r<-22$. The points denote the
results from different $L_{box}$ simulations, and the lines
represent the analytical predictions of the halo models for
different halo mass functions.} \label{fig:png}
\end{figure}

In Figure \ref{fig:png}, we show the ratio of the galaxy power
spectrum $P_g(k)$ to the linear dark matter power spectrum $P_L(k)$
only for the galaxy samples of $-23<M_r<-22$. The points denote the
results of the simulations (the shot-noise effect has been
corrected), and the different lines represent the halo model
predictions for different halo mass functions. Interestingly, the
model predictions agree very well with themselves on large scales,
indicating that the predictions on scales larger than that of
$k=0.2\mpchii$ are robust against the changes of the mass function
and corresponding bias functions. However, the halo model
predictions are larger than those of the simulations by about $16\%$
(from $0.03\mpchii$ to $0.15\mpchii$) on large scales. The wiggles
shown on large scales of the simulation results are caused by the
Baryonic Acoustic Oscillations (BAO). In the halo model
configurations, BAO is indeed embedded in the linear power spectrum
$P_L(k)$. Then the ratio $P_g(k)/P_L(k)$ for the halo model will
display no trace of oscillation on large scales, as shown in Figure
\ref{fig:png}. Since galaxies as well as the underlying dark matter
evolve in a nonlinear pattern, the resulting nonlinear power
spectrum $P_g(k)$ would, however, still remain some features of BAO,
but the nonlinear processes do suppress the BAO features. (Because
$P_L(k)$ appears in the denominator of the ratio, the peaks shown in
Figure \ref{fig:png} should actually correspond to the troughs in
the linear theory.) The differences of the simulation results
between $L1200$ and $L600$ at $k>0.7\mpchii$ are due to the limited
resolution of $L1200$.

\begin{figure*}[t]
\plotone{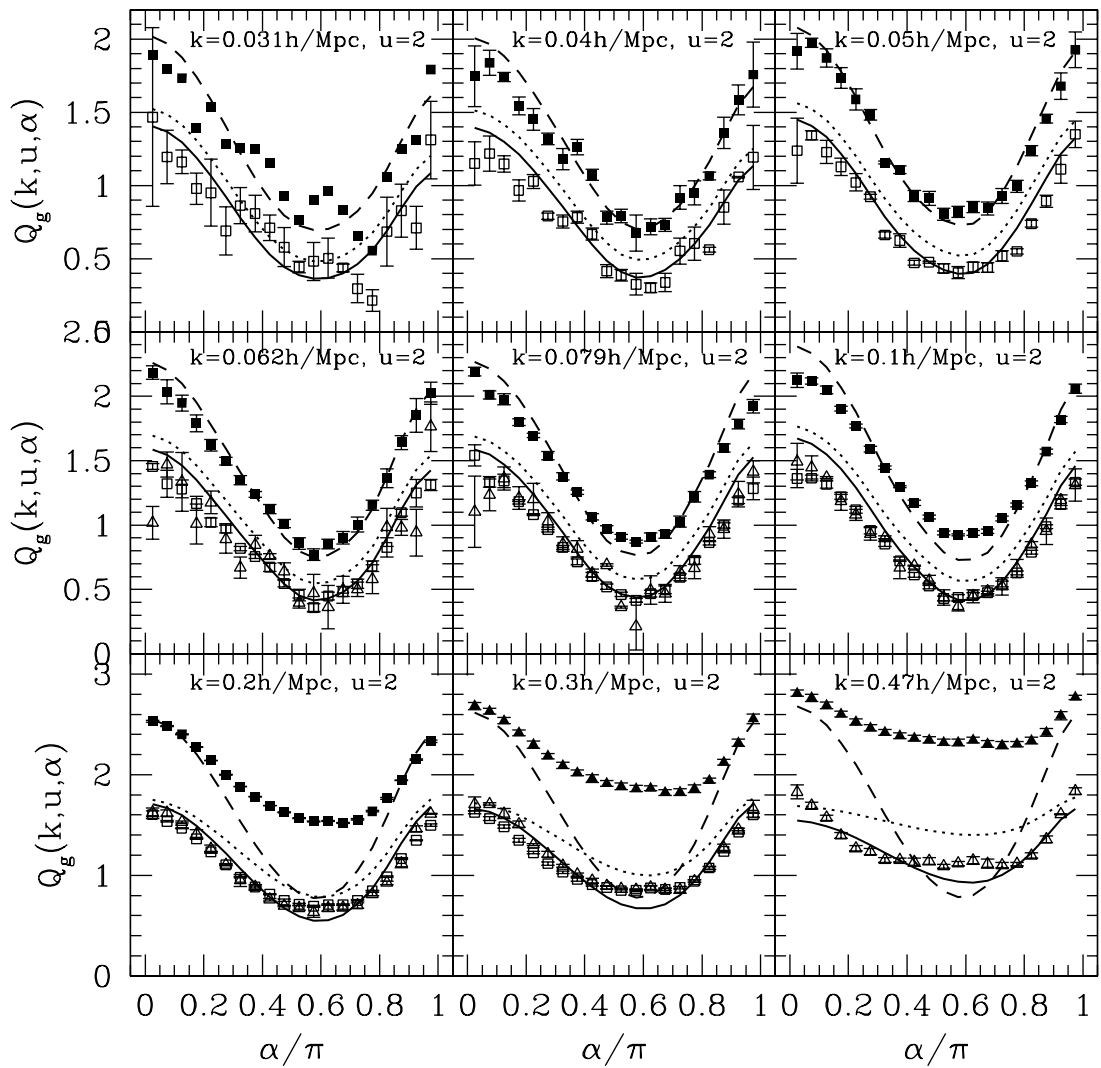} \caption{Scale and shape dependence of the galaxy
bispectrum $Q_g$ from the analytical predictions of halo models,
compared with the results from the simulations for the galaxy
samples of $-23<M_r<-22$. The square and triangle open points stand
for the results of $L1200$ and $L600$, respectively. The solid and
dotted lines represent the model predictions for PS and ST MFs,
respectively. We also show for comparison the dark matter reduced
bispectrum $Q_m$(solid symbols) and PT2 predictions(dashed lines).}
\label{fig:qg}
\end{figure*}

In halo model, $P_g(k)$ consists of two components, the 1-halo and
2-halo terms. On small scales, the 1-halo term which is actually
dominant has no dependence on the halo bias parameters. So the
differences of the halo models on the small scales mostly come from
the choices of different MFs. Our results do favor the ST MF over
the PS MF as expected. On large scales, the dominant component is
the 2-halo term and the linear halo bias factor $b_1(m)$ influences
the final predictions. The poor agreement between the simulation
results and the predictions of the PS and ST models, implies that
the MFs and the halo bias functions derived from the peak-background
split assumption, might not be fully consistent with the simulation
results. And the similar predictions between PS and ST models on
large scales also suggest that the inconsistency is probably due to
intrinsic defects of the halo model configurations. The deficiency
of halo models is also revealed by the fact that the BAO wiggles on
large scales can not be well described by the halo models. Here, we
do not consider the exclusion effect and the halo boundary effect
(see e.g., \citealt{Takada03,Smith06}), which are important only on
the quasilinear and nonlinear regimes.

In Figure \ref{fig:qg}, we show the halo model predictions of the
galaxy reduced bispectrum $Q_g(k,u,\alpha)$ where $k$ (referring to
$k_1$ in $Q_g(k_1,k_2,k_3)$) describes the size of triangle,
$u\equiv k_2/k_1$, and $\alpha\equiv arccos(\k_1 \cdot \k_2)/k_1
k_2$. The square and triangle open points stand for the simulation
results of $L1200$ and $L600$, respectively. The solid and dotted
lines represent the halo model predictions for the PS and ST MFs,
respectively. As in Figure \ref{fig:png}, the galaxies analyzed are
in the luminosity range of $-23<M_r<-22$. We also show for
comparison the dark matter reduced bispectrum $Q_m$(solid symbols)
and PT2 predictions(dashed lines). At first glance, the linear
relation of Equation (\ref{eqn:bfit}) seems to be true, confirming
our conclusions above.

On all the scales, the PS model seems to fit the simulations better,
while the ST model predicts a larger $Q_g$. The difference in the
model predictions might reflect the fact that the ST MF has more
massive halos than the PS MF. On large scales of $k<0.05\mpchii$,
the predictions of PS model are consistent with the simulation
results fairly well. But the model predictions for the ST MF are
still larger than those of the simulations except for the case of
$k=0.03\mpchii$, where the fluctuation among the simulation
realizations is dramatic.

On the intermediate scales of $0.06\mpchii \sim 0.1\mpchii$, the PS
model agrees with the simulations for isosceles triangle
configurations but is larger for the collinear configurations while
the ST model predictions are always larger than the simulation
results. On smaller scales, $Q_g$ from simulations shows a
flattening trend around the isosceles triangle configurations, as in
the bottom panels of Figure \ref{fig:qg}. But such a feature is not
seen so evidently in the halo models.

As in the case of galaxy power spectrum, the halo model is still not
accurate enough to fulfil the requirement of precision studies of
the galaxy bispectrum. Since $Q_g$ is the ratio of bispectrum $B_g$
to a sum of power spectrum products, any better agreement achieved
with the PS MF is more likely to be a coincidence, considering the
poor agreement in the $P_g(k)$ predictions. Therefore, the framework
of the halo model, including its important ingredients, needs to be
further digested in future studies.

\section{Conclusions}
\label{conclusions}

We use a set of numerical $N$-body simulations to study the large
scale behavior of the galaxy bias parameters with the bispectrum
method. We first determine the dark matter distribution from our
simulations and then construct our mock galaxy catalogs from the
semi-analytical model of \citet{Croton06}. The galaxy bias
parameters $b_1$ and $b_2$ can be simply obtained by fitting the
relation of Equation (\ref{eqn:bfit}) given the knowledge of the
dark matter and galaxy bispectra.

We find that on large scales down to $k\approx0.15\mpchii$, the bias
factors $b_1$ and $b_2$ are nearly constant. More importantly, $b_1$
obtained from the bispectrum method is consistent with that from the
power spectrum ratio (Equation (\ref{eqn:b1p})), indicating that the
linear bias factor $b_1$ can be obtained from the distribution of
galaxies. Also in general, the nonlinear bias $b_2$ is not zero; it
is of a negative value for galaxies of typical luminosity, and
increases to a positive value for the most luminous galaxies. On
scales of $k>0.15\mpchii$, the bias factors rapidly deviate from the
constancy shown on large scales. This is because the nonlinear
effects become important on these quasilinear and nonlinear regimes.
On these scales, the simple relation between $Q_g$ and $Q_m$ in
Equation (\ref{eqn:bfit}) or the assumption that $b_1$ and $b_2$ are
constant is not valid. It is therefore not effective to use this
relation to obtain the bias factors on small scales.

We also try PT2 to estimate the dark matter bispectrum in the bias
Equation (\ref{eqn:bfit}). But the poor agreement with the known
linear bias factor shows that it is very important to accurately
estimate the dark matter bispectrum when using Equation
(\ref{eqn:bfit}). This indicates that PT2 is still not accurate
enough for the dark matter bispectrum estimation even on fairly
large scales. Thus, it again implies that higher-order corrections,
such as the one-loop correction\citep{Scoccimarro96} in the
perturbation theory, are necessary for more precise estimation of
the bispectrum in the further studies.

Because of the simple structure of the halo model method, we also
incorporate HOD from our mock galaxy catalogs to model the galaxy
power spectrum and bispectrum, as shown in Figure \ref{fig:png} and
Figure \ref{fig:qg}. Overall, the analytical halo models are not
accurate enough to describe the power spectrum and bispectrum of
galaxies on the requirement of current precision studies. More work
is needed to improve the analytical halo model.

\acknowledgments

We thank the anonymous referee for helpful suggestions. This work is
supported by NSFC (10533030, 10821302, 10873028, 10878001), by the
Knowledge Innovation Program of CAS (no. KJCX2-YW-T05), and by the
973 Program (no.2007CB815402). The Millennium Run simulation used in
this paper was carried out by the Virgo Supercomputing Consortium at
the Computing Centre of the Max-Planck Society in Garching. The
semi-analytic galaxy catalogue is publicly available at
http://www.mpa-garching.mpg.de/galform/agnpaper.

\end{document}